\documentclass[aps, prl, twocolumn, groupedaddress]{revtex4}
\usepackage{bbm}
\usepackage{mathrsfs}
\usepackage{amsfonts}
\usepackage{amsmath}
\usepackage{times}
\usepackage{graphics}
\usepackage{bm}
\begin{document}
\title{Equivalence of O(3) nonlinear $\sigma$ model and the CP$^1$ model: A path integral approach}
\author{Ran Cheng, Qian Niu}
\affiliation{Department of Physics, University of Texas at Austin, Austin, TX 78712 USA}
\begin{abstract}
A rigorous proof is given on the equivalence of the O(3) nonlinear $\sigma$ model and the CP$^1$ model via path integral approach.
\end{abstract}
\maketitle

The low energy dynamics of anti-ferromagnetically correlated spins can be well described by the O(3) nonlinear $\sigma$ model (NL$\sigma$M), which abandoned the strong requirement of global order and only assumes the \emph{local} Neel order~\cite{ref:Haldane, ref:Sachdev}. The central idea is to express the local spin field $\mathbf{s}_i$ by a unimodular Neel field $|\hat{n}(x_\mu)|=1$ and a small canting field $|\hat{m}(x_\mu)|\ll 1$,
\begin{align}
\mathbf{s}_i = (-1)^i \hat{n}(x_\mu)\sqrt{1-\hat{m}(x_\mu)^2}+\hat{m}(x_\mu)\notag
\end{align}
where $x_\mu$ represents the Euclidean space of the continuum background lattice. By integrating out the small canting field $\hat{m}(x_\mu)$ in the path integral formalism, people obtained the desired effective action:
\begin{align}
\int\! \mathscr{D}^3\hat{n}\mathscr{D}^3\hat{m}\delta(\hat{n}^2-1) e^{-S[\hat{n},\hat{m}]}=\!\int\! \mathscr{D}^3\hat{n} \delta(\hat{n}^2-1) e^{-S_{\mathrm{eff}}[\hat{n}]} \notag
\end{align}
where the original action $S[\hat{n},\hat{m}]$ is derived from the Heisenberg model~\cite{ref:Sachdev} and the effective action takes the form:
\begin{align}
S_{\mathrm{eff}}[\hat{n}]=\frac1{4g}\int\mathbbm{d}x\partial_\mu\hat{n}\cdot\partial_\mu\hat{n} \notag
\end{align}
where summation over repeated indices is assumed and $g$ is the coupling strength determined by the experiments. The integration is taken over $d+1$ dimensional Euclidean space $\int\mathbbm{d}x=\int\mathrm{d}\tau\mathrm{d}\bm{x}$.

While O(3) NL$\sigma$M manifests great usefulness in the study of antiferromagnetic systems near their critical points, people usually solve the model, however, by transforming it into the celebrated $CP^1$ model~\cite{ref:Polyakov} by the Hopf map $\hat{n}=z^\dagger\hat{\sigma}z$ ($\sigma^a=$normalized Pauli matrices) where $z$ is the $CP^1$ field. The High -$T_c$ superconductivity is a good example among these situations where the $CP^1$ model is often taken as a starting point~\cite{ref:HighTc}. A striking property of the $CP^1$ model is the gauge field minimally coupled to $CP^1$ field acquires Maxwell dynamics in the long wave length limit, by which electrons with opposite spins become attractive~\cite{ref:Polyakov, ref:HighTc}. However, although the equivalence of the two models serves as a crucial foundation in these applications, a proof of their exact equivalence still bears mathematical restrictions and complexities~\cite{ref:Auerbach, ref:Banerjee}.\\

In this notes, we perform a simple but rigid proof of this equivalence via the path integral approach. To start with, we write down explicitly the amplitude for the O(3) NL$\sigma$M:
\begin{align}
\mathcal{Z}_1=\int \mathscr{D}^3 \hat{n}\delta(\hat{n}^2-1) e^{-\frac1{4g}\int\mathbbm{d}x\partial_\mu\hat{n}\cdot\partial_\mu\hat{n}} \label{eq:AmpNLSM}
\end{align}
and the amplitude for the $CP^1$ model:
\begin{align}
\mathcal{Z}_2=&\int \mathscr{D}^4z\mathscr{D}A_\mu \delta(|z|^2-1) e^{-\frac1g\int\mathbbm{d}x|(\partial_\mu-iA_\mu)z|^2} \label{eq:AmpCP1}
\end{align}
Proof of the equivalence between the two models is nothing but to show $\mathcal{Z}_1$ is proportional to $\mathcal{Z}_2$ under the Hopf map $\hat{n}=z^\dagger\hat{\sigma}z$. Let us express the $CP^1$ field as $z=(z_1, z_2)^T=(r e^{i\alpha}, s e^{i\beta})^T$, and it is easy to check that $r^2+s^2=1$ due to the constraint $\hat{n}^2=|z|^2=|z_1|^2+|z_2|^2=1$. This means the $CP^1$ field is constrained on the unit complex sphere. In terms of $r, s, \alpha,$ and $\beta$, the action in $\mathcal{Z}_1$ can be written as:
\begin{align}
\frac1{4g}&\int\mathbbm{d}x\partial_\mu\hat{n}\cdot\partial_\mu\hat{n} \notag \\ &=\frac1g\int\mathbbm{d}x[r^2s^2(\partial_\mu\alpha-\partial_\mu\beta)^2+(\partial_\mu r)^2+(\partial_\mu s)^2] \label{eq:action}
\end{align}
Next, we integrate out the gauge field in the amplitude $\mathcal{Z}_2$ which is a Gaussian integral and then express the action also in terms of the new variables $r, s, \alpha,$ and $\beta$:
\begin{align}
\mathcal{Z}_2=&\int \mathscr{D}^4z\mathscr{D}A_\mu \delta(|z|^2-1) e^{-\frac1g\int\mathbbm{d}x|(\partial_\mu-iA_\mu)z|^2} \notag\\
=&\int \mathscr{D}^4z\mathscr{D}A_\mu\delta(|z|^2-1) \notag\\
&\ \ \ e^{-\frac1g\int\mathbbm{d}x\partial_\mu z^\dagger \partial_\mu z}e^{-\frac1g\int\mathrm{d}x[A_\mu^2+iA_\mu(z^\dagger\partial_\mu z - z\partial_\mu z^\dagger)]} \notag\\
=&(\pi g)^2\int \mathscr{D}^4z\delta(|z|^2-1) e^{\frac1g\int\mathrm{d}x(r^2\partial_\mu\alpha+s^2\partial_\mu\beta)^2} \notag\\
&\qquad e^{-\frac1g\int\mathrm{d}x[r^2(\partial_\mu\alpha)^2+s^2(\partial_\mu\beta)^2+(\partial_\mu r)^2+(\partial_\mu s)^2]}
\end{align}
Considering $r^4=r^2(1-s^2)$ and $s^4=s^2(1-r^2)$, it is straightforward to show that the action in the above path integral just equals the action obtained in Eq.~\eqref{eq:action}. Put it another way, while the two path integrals have different variables, their integrands (the actions) are equal:
\begin{align}
\mathcal{Z}_1&=\int \mathscr{D}^3 \hat{n}\delta(\hat{n}^2-1) e^{-S_1[\hat{n}]} \label{eq:NLSM}\\
\mathcal{Z}_2&=(\pi g)^2\int \mathscr{D}^4z \delta(|z|^2-1) e^{-S_2[\hat{n}(z)]} \label{eq:CP1} \\
&\mbox{with}\ \ S_1[\hat{n}]=S_2[\hat{n}(z)]=S[r,s,\alpha,\beta] \notag
\end{align}
To proceed, we are to show that the entire amplitudes of Eq.~\eqref{eq:NLSM} and Eq.~\eqref{eq:CP1} are proportional, i.e., the equality:
\begin{align}
\int \!\!\mathscr{D}^4z \delta(|z|^2-1) e^{-S[\hat{n}(z)]} = c \int \!\!\mathscr{D}^3\hat{n} \delta(\hat{n}^2-1) e^{-S[\hat{n}]} \label{eq:tobeprove}
\end{align}
where $c$ is an overall constant that can be eliminated by proper normalization.

By virtue of the selection rule of the $\delta$ function and the Hopf map $\hat{n}=z^\dagger\hat{\sigma}z$ we used above, we are able to rewrite the left hand side of Eq.~\eqref{eq:tobeprove} in the form:
\begin{align}
\int \mathscr{D}^4z& \delta(|z|^2-1) e^{-S[\hat{n}(z)]} \notag\\
=&\int \mathscr{D}^4z \delta(|z|^2-1) \int \mathscr{D}^3\hat{n} \delta^3(\hat{n}-z^\dagger\hat{\sigma}z) e^{-S[\hat{n}]} \label{eq:delta}
\end{align}
thus the equality of Eq.~\eqref{eq:tobeprove} would be proved if we can show the following relation:
\begin{align}
\int \mathscr{D}^4z \delta^3(\hat{n}-z^\dagger\hat{\sigma}z)\delta(|z|^2-1)=c\ \delta(\hat{n}^2-1) \label{eq:actualprove}
\end{align}
In other words, the proof of the equivalence between the two models is now a matter of demonstrating Eq.~\eqref{eq:actualprove}.
To prove Eq.~\eqref{eq:actualprove}, we first clarify the meaning of $\mathscr{D}^4z$ by:
\begin{align}
\mathscr{D}^4z = \prod_{x_{\mu}, j=1,2}\mathrm{dRe}z_j(x_\mu)\mathrm{dIm}z_j(x_\mu)
\end{align}
and then we carry out the integral in the $r,s,\alpha,\beta$ coordinates. Since $\mathrm{Re} z_1=r\cos\alpha$, $\mathrm{Im} z_1=r\sin\alpha$, $\mathrm{Re} z_2=s\cos\beta$, and $\mathrm{Im} z_2=s\sin\beta$, the Jacobian of the coordinate transformation reads:
\begin{align}
J=\frac{\partial(\mathrm{Re} z_1, \mathrm{Im} z_1, \mathrm{Re} z_2, \mathrm{Im} z_2)}{\partial(r, \alpha, s, \beta)}=rs
\end{align}
Then the left hand side of Eq.~\eqref{eq:actualprove} becomes:
\begin{align}
\mbox{L.H.S.}=&\int_0^\infty\!rdr \int_0^\infty\!sds \int_0^{2\pi}\!\!d\alpha \int_0^{2\pi}\!\!d\beta\ \delta(r^2+s^2-1) \notag\\
&\delta(n_x-2rs\cos(\alpha-\beta))\delta(n_y+2rs\sin(\alpha-\beta)) \notag\\
&\qquad\qquad\qquad\delta(n_z-(r^2-s^2)) \notag\\
=&\frac1{16} \int_0^\infty\!dR \int_0^\infty\!dS \int_0^{4\pi}\!\!d\theta \int_{-2\pi}^{2\pi}\!\!d\phi \delta(R+S-1) \notag\\
&\delta(n_x-2\sqrt{RS}\cos(\phi))\delta(n_y+2\sqrt{RS}\sin(\phi)) \notag\\
&\qquad\qquad\qquad\delta(n_z-(R-S))
\end{align}
where some simple transformations of variables have been used. Integrating out $dR$ and $d\theta$ first and then $dS$, we obtain:
\begin{align}
\mbox{L.H.S.}&=\!\frac\pi{4} \int_0^\infty\!\!\!dS \int_{-2\pi}^{2\pi}\!\!\!d\phi\delta(n_x-2\sqrt{(1-S)S}\cos(\phi)) \notag\\
&\quad\delta(n_y+2\sqrt{(1-S)S}\sin(\phi))\delta(n_z-(1-2S)) \notag\\
&=\!\frac\pi{4}\int_{-\pi}^{\pi}\!\!\!d\phi\ \delta(n_x-\sqrt{1-{n_z}^2}\cos(\phi)) \notag\\
&\qquad\qquad\qquad\delta(n_y+\sqrt{1-{n_z}^2}\sin(\phi))
\end{align}
where in the last line we have taken into account the periodicity of the integrand so that $\int_{-2\pi}^{2\pi}=2\int_{-\pi}^{\pi}$. The last integration over $d\phi$ is somewhat tricky. Define the function $f(\phi)=\sqrt{1-{n_z}^2}\cos\phi-n_x$, it has two zero points at $\phi_0\!\!=\pm\arccos \frac{n_x}{\sqrt{1-{n_z}^2}}$, and the absolute values of its derivative at these points are:
\begin{align}
|f'(\phi_0)|=\sqrt{1-{n_z}^2}\sin\phi_0=\sqrt{1-n_x^2-n_z^2}
\end{align}
using the relation $\delta[f(\phi)]=\sum_{\phi_0}\frac{\delta(\phi-\phi_0)}{|f'(\phi_0)|}$, we obtain:
\begin{align}
\mbox{L.H.S.}&=\frac\pi{4}\int_{-\pi}^{\pi}\!\!\!d\phi\ \delta(n_y+\sqrt{1-{n_z}^2}\sin(\phi)) \notag\\
&\frac{\delta(\phi-\arccos \frac{n_x}{\sqrt{1-{n_z}^2}})+\delta(\phi+\arccos \frac{n_x}{\sqrt{1-{n_z}^2}})}{\sqrt{1-n_x^2-n_z^2}} \notag\\
&=\frac\pi2\delta(n_x^2+n_y^2+n_z^2-1) \notag\\
&=\frac\pi2\delta(\hat{n}^2-1)
\end{align}
Therefore, Eq.~\eqref{eq:actualprove} hence Eq.~\eqref{eq:tobeprove} is proved and the constant $c=\frac\pi2$. Finally, we are safe to claim the exact equivalence of the O(3) NL$\sigma$M and the $CP^1$ model in the path integral formalism:
\begin{align}
\mathcal{Z}_2&=\int \mathscr{D}^4z\mathscr{D}A_\mu \delta(|z|^2-1)\ e^{-\frac1g\int\mathrm{d}x|(\partial_\mu-iA_\mu)z|^2} \notag\\
&=\frac{\pi^3g^2}2 \int \mathscr{D}^3 \hat{n} \delta(\hat{n}^2-1) e^{-\frac1{4g}\int\mathrm{d}x\partial_\mu\hat{n}\cdot\partial_\mu\hat{n}} \notag\\
&=\frac{\pi^3g^2}2\mathcal{Z}_1
\end{align}
where the overall constant in front of $\mathcal{Z}_1$ is a trivial factor that can be eliminated by proper normalization.\\

Special thanks are offered to Xiao Li for helpful discussions and calculations. The authors are also grateful to Y. You, G. Fiete.

\end{document}